\documentclass[11pt]{article}

\usepackage{jheppub} 

\usepackage{latexsym,amsmath,amsfonts,amssymb, tensor}
\usepackage{graphicx}
\usepackage{epsf}
\usepackage{makeidx}
\usepackage{multirow}
\usepackage[all]{xy}
\usepackage{hyperref}
\usepackage{verbatim} 
\usepackage{rotating}
\usepackage{xcolor}
\usepackage{multirow}
\usepackage{bm}
\usepackage{cleveref}
\usepackage{slashed}
\usepackage[normalem]{ulem}

\usepackage{dirtytalk}

\usepackage{graphicx}
\usepackage{latexsym,amsmath,amsfonts,amssymb}
\usepackage{mathrsfs}
\usepackage{bbm}
\usepackage{bm}
\usepackage{subcaption}
\usepackage{paralist}
\usepackage{youngtab}
\usepackage{bclogo}
\usepackage{mathrsfs}

\usepackage{esvect} 

\usepackage{framed}
 
\usepackage{mathtools} 

\usepackage{tikz}
\usepackage{tikz-cd}
\usepackage{diagbox}
\usetikzlibrary{positioning}
\usetikzlibrary{calc}
\usetikzlibrary{decorations.pathreplacing,calligraphy}

\usepackage{xstring}
\usetikzlibrary{decorations.pathmorphing} 
\usetikzlibrary{decorations.markings} 
\usetikzlibrary{snakes}
\usetikzlibrary{arrows} 
\usetikzlibrary{shapes} 
\usetikzlibrary{matrix} 
\usetikzlibrary{positioning} 
\usepackage[english]{babel} 
\usepackage[autostyle]{csquotes}
\usetikzlibrary{positioning,fit}

\tikzstyle{brane}=[draw]
\tikzset{D7/.style={circle, draw=black, inner sep=0pt, fill=white, minimum size=3mm}}
\tikzset{hasse/.style={circle, fill,inner sep=2pt}}
\tikzset{flavour/.style={regular polygon,fill=white,regular polygon sides=4,inner sep=2.5pt, draw}}
\tikzset{gauge/.style={circle, draw,inner sep=2.5pt}}
\tikzset{gaugeb/.style={circle, draw,fill=black,inner sep=2.5pt}}
\tikzset{gauger/.style={circle, draw,fill=cyan,inner sep=2.5pt}}
\tikzset{gaugeg/.style={circle, draw,fill=red,inner sep=2.5pt}}
\tikzset{bd/.style={circle, draw=black, inner sep=0pt, fill=black, minimum size=2mm}}
\tikzset{wd/.style={circle, draw=black, inner sep=0pt, fill=white, minimum size=2mm}}
\tikzset{SUd/.style={circle, draw=black, inner sep=0pt, fill=yellow, minimum size=2mm}}
\tikzset{Dynkin/.style={circle, draw=black, inner sep=0pt, fill=white, minimum size=2mm}}
\tikzstyle{ligne}=[draw, thick] 
\tikzset{doublearrow/.style={ draw=black!75, color=black!75, thick, double distance=3pt, }}

\tikzset{->-/.style={decoration={
  markings,
  mark=at position #1 with {\arrow{>}}},postaction={decorate}}}
\tikzset{->>-/.style={decoration={
  markings,
  mark= between positions #1-0.05 and #1+0.05 step 0.1 with {\arrow{>}}
  },postaction={decorate}}}
\tikzset{->>>-/.style={decoration={
  markings,
  mark= between positions #1-0.1 and #1+0.1 step 0.1 with {\arrow{>}}
  },postaction={decorate}}}
  \tikzset{->>>>-/.style={decoration={
  markings,
  mark= between positions #1-0.2 and #1+0.2 step 0.1 with {\arrow{>}}
  },postaction={decorate}}}
    \tikzset{->>>>>-/.style={decoration={
  markings,
  mark= between positions #1-0.3 and #1+0.2 step 0.1 with {\arrow{>}}
  },postaction={decorate}}}
  \tikzset{->>>>>>-/.style={decoration={
  markings,
  mark= between positions #1-0.1 and #1+0.5 step 0.1 with {\arrow{>}}
  },postaction={decorate}}}
\tikzset{node3d/.style={circle,draw,minimum size=1.7cm, very thick}}
\tikzset{node1d/.style={circle,draw,minimum size=1.5cm, thick}}
\tikzset{square/.style={regular polygon,regular polygon sides=4}}
\tikzset{node3ds/.style={circle,draw, minimum size=0.8cm, very thick}}
\tikzset{node1ds/.style={circle,draw, minimum size=0.8cm, thick}}
\tikzset{squareMini/.style={regular polygon,regular polygon sides=4,draw, minimum size=1.1cm, thick}}

\tikzset{cross/.style={cross out, draw=black, minimum size=2*(#1-\pgflinewidth), inner sep=0pt, outer sep=0pt},
cross/.default={1pt}}

\numberwithin{equation}{section}

\newcommand{\be}{\begin{equation}} 
\newcommand{\ee}{\end{equation}}
\newcommand{\bea}{\begin{equation} \begin{aligned}} \newcommand{\eea}{\end{aligned} \end{equation}}

\newcommand{\bit}{\begin{itemize}} 
\newcommand{\eit}{\end{itemize}} 

\usepackage{ytableau}
\Ylinethick{0.3pt}
\Yboxdim{5pt}

\newcommand{\CN}{\mathcal{N}}

\DeclareMathAlphabet{\pazocal}{OMS}{zplm}{m}{n}

\usepackage{diagbox}
\usepackage{array}
\makeatletter
\newcommand{\thickhline}{%
    \noalign {\ifnum 0=`}\fi \hrule height 1pt
    \futurelet \reserved@a \@xhline
}
\newcolumntype{"}{@{\hskip\tabcolsep\vrule width 1pt\hskip\tabcolsep}}
\makeatother

\makeatletter
\g@addto@macro{\endtabular}{\rowfont{}}
\makeatother
\newcommand{\rowfonttype}{}
\newcommand{\rowfont}[1]{
   \gdef\rowfonttype{#1}#1%
}
\newcolumntype{L}{>{\rowfonttype}l}

\usepackage{multirow}
\usepackage{arydshln}

\begin{document}

\baselineskip=18pt  
\numberwithin{equation}{section}  
\allowdisplaybreaks  


%
%


\thispagestyle{empty}

\vspace*{0.8cm} 
\begin{center}
{{\Huge  New results on 3d $\mathcal{N}=2$ SQCD and its 3d GLSM interpretation
}}

 \vspace*{1.5cm}
Cyril Closset,  Osama Khlaif

 \vspace*{0.7cm} 

 { School of Mathematics, University of Birmingham,\\ 
Watson Building, Edgbaston, Birmingham B15 2TT, United Kingdom}\\

\vspace*{0.8cm}
\end{center}
\vspace*{.5cm}

\noindent
In this note, we review some new results we recently obtained about the infrared physics of 3d $\mathcal{N}=2$ SQCD with a unitary gauge group, in particular in the presence of a non-zero Fayet-Iliopoulos parameter and with generic values of the Chern-Simons levels.  We review the 3d GLSM (also known as 3d A-model) approach to the computation of the 3d $\mathcal{N}=2$ twisted chiral ring of half-BPS lines. For particular values of the Chern-Simons levels, this twisted chiral ring has a neat interpretation in terms of the quantum K-theory (QK) of the Grassmannian manifold. We propose a new set of line defects of the 3d gauge theory, dubbed Grothendieck lines, which represent equivariant Schubert classes in the QK ring. In particular, we show that double Grothendieck polynomials, which represent the equivariant Chern characters of the Schubert classes, arise physically as Witten indices of certain quiver supersymmetric quantum mechanics. We also explain two distinct ways how to compute K-theoretic enumerative invariants using the 3d GLSM approach. {\it This review article is a contribution to the proceedings of the GLSM@30 conference, which was held in May 2023 at the Simons Center for Geometry and Physics.}


%
\newpage

\tableofcontents


\section{Introduction}

Three-dimensional supersymmetric field theories form a rich arena for research into strongly-coupled quantum field theories. They are somewhat simpler than their four-dimensional cousins, but still allow for very interesting strongly-coupled dynamics. 3d $\mathcal{N}=2$ supersymmetric theories were first studied systematically in the late 1990's~\cite{Aharony:1997bx, Dorey:1999rb}, and then again in much detail over the last 15 years -- seminal works include Refs~\cite{Gaiotto:2007qi, Aharony:2008ug}. Of particular interest in 3d is the possibility of adding Chern-Simons interactions for the gauge fields.

Any 3d gauge theory is ultraviolet free, and flows to strong coupling in the infrared (IR). One expects that most 3d $\CN=2$ supersymmetric gauge theories flow to SCFTs in the IR. These theories can also have a vacuum moduli space, which typically consists of Higgs and Coulomb branches that intersect at the fixed point. If we consider theories with unitary gauge group, we have also the possibility of turning on Fayet-Iliopoulos (FI) parameters, in which case the vacuum moduli space can include compact Higgs branches.~\cite{Dorey:1999rb}

In this report, we summarise a number of recent results in the study of the unitary SQCD theory denoted by SQCD$[N_c, k, l, n_f, n_a]$. 
 This is a 3d $\mathcal{N}=2$ supersymmetric Yang-Mills-Chern-Simons theory with a gauge group $U(N_c)_{k,k+lN_c}$ coupled to $n_f$ and $n_a$ chiral multiplets transforming in the fundamental representation and in the antifundamental representation, respectively. Here, the Chern-Simons (CS) levels $k$ and $l$ are such that:
\begin{equation}\label{UNc def}
	U(N_c)_{k, k+lN_c} = \frac{SU(N_c)_k\times U(1)_{N_c(k+ l N_c)}}{\mathbb{Z}_{N_c}}~.
\end{equation}	
Moreover,  we need to take $k+\frac{n_f+n_a}{2} \in \mathbb{Z}$ due to 3d parity anomaly. 
This SQCD theory had been mostly studied in the case $l=0$. Recently, we gave a detailed account of its vacuum moduli space in the presence of an FI parameter~\cite{Closset:2023jiq}. In particular, we computed its flavoured Witten index~\cite{Intriligator:2013lca} for all values of the parameters. In a closely related work, we also revisited and clarified various infrared dual descriptions~\cite{Closset:2023vos}, following a number of previous works.~\cite{Aharony:1997gp, Benini:2011mf,Nii:2020ikd,Amariti:2021snj}

Here, we focus on the case $n_a=0$. Then, for a positive FI parameter, we can have the complex Grassmannian manifold $X={\rm Gr}(N_c, n_f)$ as a compact Higgs branch, in which case the low-energy physics probes interesting geometrical properties of $X$. Indeed, by compactifying the 3d $\CN=2$ gauge theory on a finite-size circle, we can think of this theory as a 2d $\CN=(2,2)$ supersymmetric gauged linear sigma model (GLSM) into $X$~\cite{Witten:1993yc}, albeit coupled to additional Kaluza-Klein (KK) modes arising from the third direction. While a strictly 2d GLSM computes the quantum cohomology of the target space $X$, this kind of {\it 3d GLSM} naturally encodes the quantum K-theory of $X$~\cite{Kapustin:2013hpk, Jockers:2018sfl}.

The quantum K-theory (QK) ring of the Grassmannian was well studied in the mathematical literature~\cite{10.1215/00127094-2010-218}, and a detailed physics perspective was presented in Refs~\cite{Jockers:2019lwe,Ueda:2019qhg, Gu:2020zpg}. This physics perspective approached the QK ring in terms of the algebra of Wilson lines wrapping the circle. In our more recent work~\cite{Closset:2023bdr}, we revisited the QK/3d GLSM correspondence and we presented a new basis of defect line operators, dubbed Grothendieck lines, which directly and naturally reproduce the mathematical basis of the QK ring given in terms of structure sheaves of Schubert varieties in $X$. These defect lines are defined in the UV by coupling the 3d SQCD theory to a 1d unitary gauge theory living on the line. In this short note, we review this construction. We also explain the general 3d A-model perspective~\cite{Closset:2017zgf, Closset:2019hyt}, which is a 2d effective field theory approach to the 3d GLSM. Finally, we explain how to efficiently compute the quantum K-theory invariants in terms of topologically twisted indices -- that is,  using the 3d $\CN=2$ partitions on $S^2\times S^1$, which admit well-understood exact expressions~\cite{Nekrasov:2014xaa, Gukov:2015sna, Benini:2015noa, Benini:2016hjo, Closset:2016arn}.

\medskip
\noindent
This note is organised as follows. In section~\ref{sec:3dAmod}, we review the 3d A-model perspective. In section~\ref{sec:mod-space-vacaua}, we present our new results for the moduli space of SQCD with $n_a=0$.  In section~\ref{sec:GrLines}, we explain how to construct the Grothendieck lines and how to compute QK invariants of $X$ using the 3d A-model.

\section{The 3d A-model into the Grassmannian}\label{sec:3dAmod}

Let us start with 3d $\mathcal{N}=2$ SQCD$[N_c, k, l, n_f, 0]$, with gauge group~\eqref{UNc def}. One interesting feature of these theories is the structure of their moduli space of vacua $\mathcal{M}[N_c, k, l, n_f]$ which, for fixed gauge-group rank $N_c$ and number of fundamentals $n_f$, depends in an intricate way on the CS levels $k$ and $l$~\cite{Intriligator:2013lca, Closset:2023jiq}. As we will review in the next section, for a positive 3d real FI parameter $\xi$, there exists a particular set of values for $(k,l)$ such that the moduli space of the theory consists of a single Higgs branch, the complex Grassmannian manifold:
\begin{equation}
	\mathcal{M}_{\text{Higgs}} = \text{Gr}(N_c, n_f)~.
\end{equation}
We refer to this set of values of $(k,l)$ as the geometric window.

\subsection{Putting the theory on $\Sigma_g\times S^1_{\beta}$}
To make the geometric interpretation of our 3d SQCD more apparent, we put the theory on $\Sigma_g\times S^1_{\beta}$ with $\Sigma_g$ being a compact closed genus-$g$ Riemann surface along which we perform a topological A-twist. Taking the CS levels $k$ and $l$ to be in the geometric window,  the 3d theory becomes a GLSM that flows in the infrared to a 3d NLSM with the target being the Grassmannian variety:
\begin{equation}
\text{3d NLSM} \; : \; \Sigma_g \times S_\beta^1 \longrightarrow X \equiv \text{Gr}(N_c, n_f)~.
\end{equation}
From this perspective, we can view our 3d theory as a 2d $\mathcal{N}=(2,2)$ supersymmetric gauge theory, A-twisted along $\Sigma_g$, coupled to an infinite number of Kaluza-Klein (KK) modes arising from the  $S^1_\beta$ compactification. The low-energy physics is conveniently described in terms of the semi-classical Coulomb branch of the theory,  wherein the dynamics of the 2d abelian vector multiplets (for the maximal torus $\prod_{a=1}^{N_c}U(1)_a$ of $U(N_c)$) is controlled by the twisted superpotential $\mathcal{W}(u,v, \tau)$. The latter takes the explicit, one-loop exact form~\cite{Closset:2019hyt}:
\begin{multline}\label{W-full}
	\mathcal{W}(u, v, \tau) =   \frac{1}{(2\pi i)^2} \sum_{\alpha=1}^{n_f} \sum_{a=1}^{N_c}  \text{Li}_2(x_a y_\alpha^{-1})
+ \tau \sum_{a=1}^{N_c} u_a \\
   +\frac{k+\frac{n_f}{2}}{2}\sum_{a=1}^{N_c} u_a(u_a+1) + \frac{l}{2} \left(\left(\sum_{a=1}^{N_c} u_a\right)^2 +\sum_{a=1}^{N_c} u_a \right)~,
\end{multline}
where we included the one-loop contribution of KK towers of fields from the massive chiral multiplets.  
Here,  $u_a$ are the Coulomb branch parameters (complex scalars, made dimensionless using the KK scale, whose imaginary part are the 3d real scalars $\sigma_a$), and  $\tau$ is the complexified  FI parameter in 2d, whose imaginary part is the real 3d FI parameter $\xi$. 

In curved space, one also needs to consider the effective dilaton potential $\Omega(u,v)$, which couples to the the curvature of $\Sigma_g$ and which takes the explicit form~\cite{Nekrasov:2014xaa, Closset:2019hyt}:
\begin{equation}\label{Omega-full}
	e^{2\pi i \Omega} = \prod_{\alpha=1}^{n_f}\prod_{a=1}^{N_c} (1- x_a y^{-1}_\alpha)^{-r+1} \prod_{\substack{a,b\\ a\neq b}} (1- x_a x_b^{-1})^{-1}~,
\end{equation}
where $r$ is the charge of the matter multiplets under the $U(1)_R$ R-symmetry. Here and in \eqref{W-full}, we introduced the following parameters:
\begin{align}
\begin{split}
	x_a \equiv e^{2\pi i u_a } \sim e^{-2\pi \sigma_a}~, \qquad a = 1,  \cdots, N_c~,\\
	y_\alpha \equiv e^{2\pi i v_a}  \sim e^{-2\pi m_{\alpha}}~, \qquad \alpha = 1, \cdots, n_f~,
\end{split}
\end{align}
where $v_a$ are `equivariant parameters' (twisted masses) for the $SU(n_f)$ flavour symmetry, with $\prod_{\alpha=1}^{n_f} y_\alpha=1$. Moreover, we use the shorthand notation $\text{det\;}x = \prod_{a=1}^{N_c} x_a$.

Using the A-model approach, one can show that the 3d twisted index on a genus-$g$ Riemann surface is given by the following explicit expression~\cite{Nekrasov:2014xaa}:
\begin{equation}\label{genus-g-1}
	Z_{\Sigma_g\times S^1_{\beta}} = \sum_{\hat{x}\in \mathcal{S}_{\text{BE}}} \mathcal{H}^{g-1}(\hat{x})~,
\end{equation}
where the set $\mathcal{S}_{\text{BE}}$ is the set of Bethe vacua (BV), which are defined as the solutions to the so-called Bethe ansatz equations (BAEs):
\begin{equation}\label{BE}
\mathcal{S}_{\text{BE}} = \Big\{\; \hat{x} \;  \; \Big| \;e^{2\pi i \frac{\partial\mathcal{W}}{\partial u_a}}\mid_{\hat{x}} =1~, \, \forall a\; \;\; {\text{ and}} \;\; \hat{x}_{a}\neq \hat{x}_{b}~, \forall a\neq b\, \Big\}\Big/S_{N_c}~.
\end{equation}
Here, the so-called handle-gluing operator $\mathcal{H}$ is given by:
\begin{equation}
	\mathcal{H}(u, v, \tau) = e^{2\pi i \Omega} \;\det_{1\leq a, b\leq N_c} \left(\frac{\partial^2\mathcal{W}}{\partial u_a \partial u_b}\right)~.
\end{equation}

\subsection{Frobenius algebra and Gr\"obner basis algorithm}\label{subsec:frobalg}
The explicit formula for the 3d twisted index given in \eqref{genus-g-1} can be generalized to the case where we include half-BPS line operators, for instance supersymmetric Wilson lines, wrapping the $S^1$ factor and sitting at some points $p\in \Sigma_g$. More explicitly, taking 
\begin{equation}
    \mathcal{L}_1(p_1), \cdots, \mathcal{L}_n(p_n) \in \mathbb{Z}(q,y)[x_1, \cdots, x_{N_c}]~,
\end{equation}
where the lines are represented by certain $U(N_c)$ characters (symmetric polynomials in $x_a$), 
one can show that the genus-$g$ correlation function can be written as:
\begin{equation}\label{n-gneus-g-corr}
\left<\mathcal{L}_1(p_1)\cdots\mathcal{L}_n(p_n)\right>_{\Sigma_g\times S_\beta^1} = \sum_{\hat{x}\in \mathcal{S}_{\text{BE}}} \mathcal{H}(\hat{x})^{g-1} \prod_{s=1}^n \mathcal{L}_{s}(\hat{x})~.
\end{equation} 

From the point of view of the Riemann surface, we have a 2d TQFT (obtained from the $A$-twist of the effective 2d $\CN=(2,2)$ gauge theory) which is then associated with a Frobenius algebra. Physically, this is the 3d twisted chiral ring $\mathcal{R}^{\text{3d}}$ of half-BPS line operators wrapping the circle. Choosing a basis for the half-BPS line operators of the theory $\{\mathcal{L}_\mu\}$, with $\mu$ some index, we define the nonsingular Frobenius metric as follows:
\begin{equation}\label{top-metric}
	\eta_{\mu\nu} (\mathcal{L}) := \left<\mathcal{L}_\mu \; \mathcal{L}_\nu\right>_{\mathbb{P}^1\times S^1_{\beta}}~,
\end{equation} 
where we made explicit the dependence of the topological metric on a specific choice of basis for the line operators in $\mathcal{R}^{\text{3d}}$, and we focus here on the genus-$0$ case. Similarly, we can define the structure constants of $\mathcal{R}^{\text{3d}}$ in terms of the 3-point functions:
\begin{equation}\label{struc-const}
	\mathcal{N}_{\mu\nu\lambda}(\mathcal{L}) = \left<\mathcal{L}_\mu\;\mathcal{L}_\nu\;\mathcal{L}_{\lambda}\right>_{\mathbb{P}^1\times S^1_{\beta}} ~.
\end{equation}

The components of the topological metric \eqref{top-metric} and the structure constants \eqref{struc-const} can be computed explicitly from supersymmetric localization using the formula \eqref{n-gneus-g-corr} above. Doing so, one can write down the ring relations for $\mathcal{R}^{\text{3d}}$:
\begin{equation}\label{ring-relations-general}
	\mathcal{L}_\mu\; \star \; \mathcal{L}_{\nu} = \sum_{\lambda} {\mathcal{N}_{\mu \nu}}^{\lambda}(\mathcal{L})\; \mathcal{L}_{\lambda}~,
\end{equation}
where we used the inverse of the topological metric, $\eta^{\mu\nu}(\mathcal{L})$, to raise one of the indices of the structure constants:
\begin{equation}
{\mathcal{N}_{\mu\nu}}^{\lambda}(\mathcal{L}) = \eta^{\lambda\rho}(\mathcal{L})\;\mathcal{N}_{\mu \nu \rho}(\mathcal{L})~.
\end{equation}

Now, let us pick the CS levels $k$ and $l$ to be in the geometric window, for some fixed $N_c$ and $n_f$. The question is: what is the geometric interpretation of the twisted chiral ring $\mathcal{R}^{\text{3d}}$? To answer this question, we recall that, in a 2d GLSM,  twisted chiral local operators represent cohomology classes of the target space $X$. In this spirit, in the 3d uplift, line operators $\mathcal{L}_\mu$ are expected to represent classes of coherent sheaves in the Grothendieck ring K$(X)$ of the target space. Therefore, the ring relations \eqref{ring-relations-general} are interpreted as the relations of some ``generalised" quantum K-theory ring of the Grassmannian variety $X = \text{Gr}(N_c, n_f)$.

For example, for the special choice of the CS levels:
\begin{equation}\label{qk-cs-levels}
	k = N_c - \frac{n_f}{2}~, \qquad l = -1~, 
\end{equation}
and taking $r=0$ for the R-charge of the matter fields, it has been established that the twisted chiral ring relations are those of the ordinary quantum K-theory of $X$: QK$(X)$~\cite{Jockers:2019lwe,Ueda:2019qhg}. (For the same price, we get the equivariant QK ring, QK$_T(X)$.) A standard basis for writing down this ring is that of the Schubert classes $[\mathcal{O}_{\mu}]\in \text{K}(X)$~\cite{10.1215/00127094-2010-218}. These are equivalence classes of the structure sheaves supported on the Schubert subvarieties $X_\mu\subset X$. In this case, the index $\mu$ denotes an $N_c$-partition whose Young tableau fits inside an $N_c\times (n_f-N_c)$ rectangle. 

Any coherent sheaf can be represented by its Chern character, which is polynomial in $x_a$ (mathematically, $\log{x_a}$ are the Chern roots of the tautological vector bundle of $X$). The (equivariant) structure sheaves $\mathcal{O}_{\mu}$ can then be represented by the (double) Grothendieck polynomials, ${\rm ch}_T(\mathcal{O}_{\mu})=\mathfrak{G}_{\mu}(x,y)$, which are given by~\cite{ikeda2013k}:
\begin{equation}\label{doub-groth-poly}
\mathfrak{G}_{\mu} (x,y) = \frac{\det_{1\leq a, b \leq N_c} \left(x^{b-1}_{a} \prod_{\alpha=1}^{\mu_b+N_c-b} \left(1-x_a y_{\alpha}^{-1}\right)\right)}{\prod_{1\leq a< b\leq N_c}(x_a - x_b)}~.
\end{equation}
These are symmetric polynomials in the Coulomb branch variables $x_a$. This property will prove useful momentarily. 

One question to ask at this point is: what is the half-BPS line operator $\mathcal{O}_\mu$ in the 3d GLSM that flows to this Schubert class in the IR? We will answer this question in section~\ref{sec:GrLines}. For now, assuming these line operators do exist, let us review how one can compute the ring structure of QK$(X)$ from the 3d A-model perspective, using the so-called Gr\"obner basis techniques~\cite{Closset:2023vos}.

\medskip
\noindent
\textbf{Gr\"obner basis algorithm.} From the 3d A-model point of view, the 3d twisted chiral ring $\mathcal{R}^{\text{3d}}$ is defined as:
\begin{equation}\label{3d-tcr}
	\mathcal{R}^{\text{3d}}  = \frac{\mathbb{Z}(q,y)[x_1, \cdots, x_{N_c}]^{S_{N_c}}}{(\partial\mathcal{W})}~,
\end{equation} 
where the relations defining the ring are none other than the BAEs that we mentioned earlier in defining the set $\mathcal{S}_{\text{BE}}$ in \eqref{BE}. Due to the residual gauge symmetry $S_{N_c}$ on the Coulomb branch, on can symmetrize the elements of the ring \eqref{3d-tcr} using any complete basis of symmetric polynomials in the variables $x_1, \cdots, x_{N_c}$. Our candidate here are none other than the double Grothendieck polynomials $\mathfrak{G}_{\mu}(x,y)$ given in \eqref{doub-groth-poly}.

To find the ring relations for $\mathcal{R}^{\text{3d}}$, we use a classical Gr\"obner basis algorithm~\cite{Closset:2023vos} to reduce the BAEs to relations between the double Grothendieck polynomials. In this way, identifying the Grothendieck polynomials with the corresponding coherent sheaves, we get the defining relations of whatever generalised QK$(X)$ we have on the geometry side of the story. To see how this works, let us define the following \textit{Bethe ideal}:
\begin{equation}
	\mathcal{I}_{\text{BE}}^{(x, w, \mathcal{O})} = (P, \hat{P}, \hat{G}, \hat{W})\subset \mathbb{Z}(q,y)[x_1, \cdots, x_{N_c}, w, \mathcal{O}]~, 
\end{equation}
where $P_a(x)$ are the polynomials that define the BAEs and in terms of which we define the symmetrized polynomials:
\begin{equation}
	\hat{P}_{ab} = \frac{P_a(x)-P_b(x)}{x_a-x_b} \in \mathbb{Z}(q,y)[x_1, \cdots, x_{N_c}]~, \qquad a>b~.
\end{equation}
Moreover, we introduced the variables $\mathcal{O}_{\mu}$ which, in $\mathcal{R}^{\text{3d}}$ we identify with $\mathfrak{G}_{\mu}(x,y)$ by taking:
\begin{equation}
	\hat{G}_{\mu}(x, \mathcal{O}_{\mu}) \equiv \mathfrak{G}_{\mu} (x,y) - \mathcal{O}_\mu~.
\end{equation}
The variable $w$ is introduced to insure that no non-phyiscal vacua is to be included in the computations via:
\begin{equation}
	\hat{W} = w \text{det\;}x - 1~.
\end{equation}

One can reduce the symmetric Bethe ideal $\mathcal{I}_{\text{BE}}^{(x, w, \mathcal{O})}$ so that it is written in terms of the symmetric Grothendieck polynomials $\mathfrak{G}_{\mu}$ only, by using the relations $\hat{G}_\mu = 0$. As a result we get the \textit{Grothendieck ideal} $\mathcal{I}_{\text{BE}}^{(\mathcal{O})}$ which can be explicitly computed using the Gr\"obner basis techniques~\cite{Closset:2023bdr}. The ring $\mathcal{R}^{\text{3d}}$, therefore, is presented explicitly as:
\begin{equation}
	\mathcal{R}^{\text{3d}} \cong \frac{\mathbb{Z}(q,y)[\mathcal{O}]}{\mathcal{I}_{\text{BE}}^{(\mathcal{O})}}~, 
\end{equation}
and the Grothendieck ideal is nothing other than the ideal generated by the ring relations:
\begin{equation}
	\mathcal{O}_{\mu} \star \mathcal{O}_\nu = \sum_{\lambda} {\mathcal{N}_{\mu\nu}}^{\lambda}\; \mathcal{O}_{\lambda}
\end{equation}
In this way, the 3d GLSM computation of $\mathcal{R}^{\text{3d}}$, with the CS levels~\eqref{3d-tcr},  reproduces exactly the equivariant QK ring of $X={\rm Gr}(N_c, n_f)$.

\medskip 
\noindent
\textbf{Equivariant QK ring of $\mathbb{P}^2$.} As a simple example, let us consider a 3d $\mathcal{N} =2$ $U(1)_{-\frac{3}{2}}$ gauge theory with $3$ chiral multiplets of charges $+1$. This a 3d GLSM with target space $\mathbb{P}^2$. Using the Gr\"obner basis algorithm discussed above, we find the following relations of QK$(\mathbb{P}^2)$:
\begin{align}\label{QKP2}
\begin{split}
& \mathcal{O}_1\star \mathcal{O}_1 =\;  \left(1-\frac{y_2}{y_1}\right) \mathcal{O}_1 + \frac{y_2}{ y_1} \mathcal{O}_2~,\\
& \mathcal{O}_1\star \mathcal{O}_2 =\;  \left(1-\frac{y_3}{y_1}\right) \mathcal{O}_2 + \frac{y_3}{y_1} q~,\\
& \mathcal{O}_2\star\mathcal{O}_2 =\;  \left(1-\frac{y_3}{y_1}\right)\left(1-\frac{y_3}{y_2}\right) \mathcal{O}_2 + \frac{y_3}{ y_2} q \mathcal{O}_1 + \left(1-\frac{y_3}{y_2}\right)\frac{y_3}{y_1} q~,\\
\end{split}
\end{align}
which precisely matches the mathematical results~\cite{10.1215/00127094-2010-218}.


\section{Moduli space of 3d vacua}\label{sec:mod-space-vacaua}
As explained above, the 3d gauge theory on $\Sigma_g\times S^1_\beta$ becomes a GLSM with a purely geometric phase for a particular set of values for the CS levels $k$ and $l$, which we referred to as the geometric window of the theory. For these values and for $\xi>0$, the moduli space of vacua of the theory in $\mathbb{R}^3$  consists only of the Higgs branch $X = \text{Gr}(N_c, n_f)$. At the level of the Witten index of the SQCD theory, these values of $k$ and $l$ correspond to the cases when the effective number of vacua ({\it i.e} the Witten index) is equal to the Euler characteristic of $X$:
\begin{equation}
\chi(\text{Gr}(N_c, n_f))= \begin{pmatrix}
    n_f\\ N_c
\end{pmatrix}~.  
\end{equation} 
In this section, we review recent results on the explicit form of the moduli space of vacua of the 3d theory for generic values of $k$ and $l$, with fixed $N_c$ and $n_f$. In particular, this computation determines the geometric window in all cases.

\subsection{Structure of the moduli space of vacua}
Recall that we have a 3d $\CN=2$ gauge theory with gauge group $U(N_c)_{k, k+lN_c}$ coupled with $n_f$ matter multiplets in the fundamental representation. Upon diagonalising the real scalar in the vector multiplet, $\sigma= {\rm diag}(\sigma_a)$, the semi-classical 3d vacuum equations read~\cite{Intriligator:2013lca}:
\begin{align}\label{vac-eqns}
\begin{split}
	&(\sigma_a - m_i)\,\phi_{i}^a =0~,  \qquad\qquad  i = 1,  {\cdots}, n_f~,  \quad a = 1, \cdots, N_c~,\\
	&\sum_{i=1}^{n_f} \phi_{a}^{\dagger i} \phi_{i}^b  = \frac{{\delta_{a}}^b}{2\pi} F_a(\sigma)~,   \qquad a, b = 1, \cdots, N_c~,
\end{split}
\end{align}
 where:
\begin{equation}
F_a(\sigma) = \xi + k \sigma_a + l \sum_{b=1}^{N_c} \sigma_b +\frac{1}{2}\sum_{i=1}^{n_f} |\sigma_a-m_i|~.
\end{equation}
The fields $\phi_i= (\phi^a_i)$ are the complex scalars of the fundamental chiral multiplets, and $m_i$ are the real masses associated with the $SU(n_f)$ flavour symmetry group. Recall also that $\xi$ is the 3d real FI parameter, which we take to be non-vanishing.

In the following, let us set the real masses $m_i = 0$. (For generic $m_i$, we  only have discrete vacua. Here we are interested in the non-trivial Higgs branches that may arise in the massless limit.) In this case, depending on the values of $k$ and $l$ and the sign of the FI parameter $\xi$, the solutions of these equations can be of the following types:

\medskip
\noindent
\textbf{Higgs vacuum.} This type of solution appears at the origin of the classical 3d  Coulomb branch: $\sigma_a = 0, \forall a$.  In this case the equations \eqref{vac-eqns} parameterise the Grassmannian variety Gr$(N_c, n_f)$ if and only if $\xi>0$.

\medskip
\noindent
\textbf{Topological vacua.} These types of vacua show up at generic points of the classical Coulomb branch; that is, points where all the components of the adjoint real scalar $\sigma$ are non vanishing. From the first equation in \eqref{vac-eqns}, one can see that these non-vanishing $\sigma$'s give masses to the chiral multiplets, hence one needs to integrate them out. At low energy, this leaves us with a topological field theory, which can be described as a $U(p)\times U(N_c-p)$ 3d $\CN=2$ Chern-Simons theory with (mixed) CS levels determined by integrating out the massive chirals. The ranks of the two gauge groups are determined by the number of $\sigma_a$'s that are positive and negative, respectively, in this solution. 

\medskip
\noindent
\textbf{Higgs-topological vacua.} Another possible form for the solutions of the vacuum equations \eqref{vac-eqns} appear at points where some of the $\sigma_a$'s are taken to be zero and the rest are non-vanishing. In this case, and depending on the sign of the FI parameter, the vacua take the form Gr$(p, n_f)\times U(N_c-p)$, where the first factor is a Higgs branch, and the second factor is a topological sector that arises from integrating out massive matter multiplets that get their masses from the non-vanishing components of $\sigma$. By explicit computation, one can show that this is the most general possible form for the hybrid vacua. In particular, one can show that solutions of the form Gr$(N_c-p-q, n_f) \times U(p)\times U(q)$ cannot appear.

\medskip
\noindent 
\textbf{Strongly-coupled vacua.} One last possibility results from the fact that we are analysing the vacuum equations in the semi-classical limit. In this limit, quantum effects that might give rise to strongly coupled vacua in the interior of the classical Coulomb branch are not taken into account. These effects conjecturally arise when a non-compact Coulomb branch direction is allowed by our semi-classical analysis. It turns out that we can always infer the contribution of these putative strongly-coupled vacua to the Witten index by various indirect ways~\cite{Closset:2023jiq}. Such vacua can only arise when $|k|=\frac{n_f}{2}$.

\subsection{An example: $U(2)_{k, k+2l}$ with $n_f=4$}
Let us present some examples for the moduli spaces of vacua that we get from different values of $k$ and $l$. For simplicity, let us fix the rank of the gauge group to be $N_c=2$ and couple the theory to $4$ fundamental matter multiplets. In this case, we find the forms of the moduli spaces of vacua shown in table \ref{tab: moduliGr24}, in two dissecting phases distinguished by the sign of the FI parameter.
\begin{table}[t] 
\renewcommand{\arraystretch}{1.1}
\centering
\begin{equation}
 \begin{array}{|c|c||c|c|}
 \hline
 k&l &~ \xi > 0 ~\text{ phase } & ~\xi<0 ~ \text{ phase } \\ \hline \hline
  0&10& \text{Gr}(2,4)\oplus~ U(2)_{-2,18}~ &~ U(2)_{2,22} \oplus U(\underbrace{1)_{12} \times U(1}_{10})_8 ~\\ \hline
 1&3& \text{Gr}(2,4) \oplus U(\underbrace{1)_6\times U(1}_3)_2 & U(2)_{3,9} \\ \hline
 3&-2&  \text{Gr}(2,4) &   \mathbb{CP}^{3} \times U(1)_{-1} \oplus U(2)_{5,1} \\ \hline
 4&7&  \text{Gr}(2,4)\oplus \mathbb{CP}^3\times U(1)_9 \oplus U(2)_{2,16} & U(2)_{6,20}\\ \hline
 5&-6&  \text{Gr}(2,4)\oplus U(2)_{7,-5} & \mathbb{CP}^{3}\times U(1)_{-3}\oplus U(2)_{3,-9} \\ \hline
 6&-4&  \text{Gr}(2,4) & U(2)_{4,-4}  \\ \hline
7&-9&  \text{Gr}(2,4) \oplus U(2)_{9,-9} & \mathbb{CP}^{n_f-1} \times U(1)_{-4} \oplus U(2)_{5,-13} \\ \hline
8&8&  \text{Gr}(2,4)\oplus \mathbb{CP}^{3}\times U(1)_{14}\oplus U(2)_{6,22} & U(2)_{10,26} \\ \hline
9&10&  \text{Gr}(2,4) \oplus \mathbb{CP}^3\times U(1)_{17} \oplus U(2)_{7,77} & U(2)_{11,31} \\ \hline
10&5&  \text{Gr}(2,4)\oplus \mathbb{CP}^3\times U(1)_{13}\oplus U(2)_{8,18} & U(2)_{12,22} \\ \hline
\end{array}
\end{equation}
\caption{Moduli spaces of vacua for $U(2)$ theory coupled with $4$ fundamental multiplets and different values of the levels $k$ and $l$. We include both phases of $\xi$, the positive and negative one.}
\label{tab: moduliGr24}
\end{table}

From the explicit knowledge of the moduli space (at least in case when no strongly-coupled vacua arise), one can compute the 3d Witten index by computing the contribution coming from each one of the components. For example, in the table \ref{tab: moduliGr24}, one can compute the index in both phases of $\xi$ and see that they do match, in agreement with the fact that we cannot have any non-trivial wall-crossing phenomena in 3d $\CN=2$ supersymmetric gauge theories~\cite{Intriligator:2013lca}.

\medskip
\noindent
\textbf{The geometric window: tentative geometric interpretation.} For fixed values of $N_c$ and $n_f$, we find several values of the CS levels $k$ and $l$ in the geometric window of the theory other than the standard ones \eqref{qk-cs-levels}. For example, we can see from table~\ref{tab: moduliGr24} that we get the Higgs branch Gr$(2,4)$ for at least two distinct choices of CS levels.

The 3d twisted chiral ring $\mathcal{R}^{\text{3d}}$ that we get for the choice \eqref{qk-cs-levels} is known to be isomorphic to QK$_{T}(X)$, with the equivariant direction $T\subset SU(n_f)$ being the maximal torus of the global symmetry.  Recently,  Ruan and Zhang introduced a so-called level structure that generalises the definition of the quantum K-theory ring of $X$~\cite{Ruan2018TheLS}, and there is some good evidence that this is realised physically by changing the CS levels away from the `standard' choice~\eqref{qk-cs-levels}~\cite{Jockers:2019lwe}.  A precise physical account of this relation is still missing, however; we shall address this point in future work.

\section{Grothendieck lines in 3d GLSM and quantum K-theory}\label{sec:GrLines}
When writing down the relations describing the structure of the (equivariant) quantum K-theory ring of the Grassmannian variety, the  most natural basis is that of the (equivariant) Schubert classes $[\mathcal{O}_\mu]$, as mentioned above. These are the classes of the structure sheaves supported on the Schubert cells $\text{C}_\mu$ that can be used to decompose the Grassmannian. In this section, we answer the question posed earlier concerning the 3d gauge theory/quantum K-theory dictionary: what are the half-BPS line operators in the 3d $\CN=2$ gauge theory that flow to these Schubert classes in the IR? 

In this section, we will first construct the sought-after line operators. Next, we will come back to our earlier discussion of the Frobenius algebra and present explicit formulas for the structure constants and the topological metric associated with the quantum K-theory ring QK$_T(\text{Gr}(N_c, n_f))$.

\subsection{Defect lines and 1d quivers}
We propose that these line operators can be constructed in the UV by coupling the 3d gauge theory to a particular 1d $\CN=2$ supersymmetric gauged quantum mechanics that does the job. Hence, let us  first elaborate a little bit more on what this ``job" is. 

The Grassmannian variety $X={\rm Gr}(N_c, n_f)$ can be decomposed into Schubert cells $\text{C}_\mu$ that are indexed by partitions $\mu$ which fit inside a $N_c\times (n_f-N_c)$ Young tableau. Moreover, the Schubert varieties (defined as the closure of the Schubert cells) generate the homology of $X$, and $\mathcal{O}_\mu$ denotes the structure sheaf of the corresponding Schubert variety. Concretely, points in the Schubert cells are $N_c$-dimensional vector subspaces of $\mathbb{C}^{n_f}$ satisfying particular conditions defined in terms of the elements of the partition $\mu$. These cells are usually represented by matrices whose rows are none other than the $\mathbb{C}^{n_f}$-vectors spanning these $N_c$-dimensional subspaces (see below for some  explicit examples). 

When computing K-theoretic Gromov-Witten invariants, the Schubert classes $[\mathcal{O}_\mu]$ appear in the guise of their (equivariant) Chern characters. Explicitly, these are given in terms of the (double) Grothendieck polynomials $\mathfrak{G}_{\mu}(x,y)$ that we introduced in \eqref{doub-groth-poly}. From this point of view, the gauge parameters $x_a\sim e^{-2\pi\beta\sigma_a}$ are interpreted as the exponential of the roots of the tautological vector bundle over Gr$(N_c, n_f)$, as already mentioned. Meanwhile, the mass parameters $y_\alpha \sim e^{-2\pi\beta m_\alpha}$ are the weights associated with the action of the isometry group $SU(n_f)$ of Gr$(N_c, n_f)$.

\medskip
\noindent
\textbf{Grothendieck lines: constraining the form of the matter matrix.} Now that we have these properties of the Schubert classes at the back of our mind, let us explain our definition of the UV line operators. For obvious reasons, we call them the Grothendieck lines.

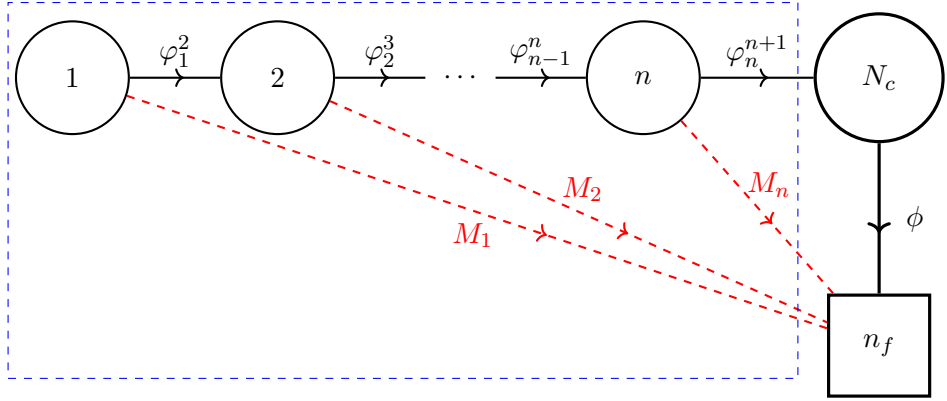
\begin{figure}[t]
    \centering
    \begin{tikzpicture}[baseline=1mm]
\node[node3d] (1) []{$N_c$};
\node[square,draw,minimum size=1.9cm,very thick] (2) [below= 2 of 1]{$n_f$};
\node[node1d] (rn) [left= 1.5 of 1]{$n$};
\node[] (inter1) [left= 1.2 of rn]{$\;\cdots\;$};
\node[node1d] (r2) [left= 1.2 of inter1]{$2$};
\node[node1d] (r1) [left= 1.2 of r2]{$1$};
\draw[->-=0.6,very thick] (1) --  (2) node[midway,right]{$\; \; \phi$};
\draw[->-=0.6, thick] (rn) --  (1) node[midway,above]{$\varphi_n^{n+1}$};
\draw[->-=0.6, thick] (inter1) --  (rn) node[midway,above]{$\varphi_{n-1}^{n}$};
\draw[->-=0.6, thick] (r2) -- (inter1) node[midway,above]{$\varphi_2^{3}$};
\draw[->-=0.6, thick] (r1) -- (r2) node[midway,above]{$\varphi_1^{2}$};
\draw[->-=0.6,red, dashed, thick] (r1) -- (2) node[midway,below]{$M_1\;$};
\draw[->-=0.6,red, dashed, thick] (r2) -- (2) node[midway,above]{$\; M_2$};
\draw[->-=0.6,red, dashed, thick] (rn) -- (2) node[midway,above]{$\;\; \; M_n$};
   \node[fit=(r1)(rn), dashed, blue, draw, inner sep=9.5pt, minimum width=10.5cm,minimum height=5cm, shift={(0.6cm,-1.5cm)}] (rn) {};
\end{tikzpicture}
    \caption{Generic Grothendieck defect $\mathcal{L}_\mu$ with $n\leq N_c$. The numbers of fermi multiplets, $M_\ell$, are given in terms of the partition $\mu$ as explained in the main text.}
    \label{fig: groth-defect}
\end{figure}
Let us consider a partition
    $\mu = [\mu_1, \mu_2, \cdots, \mu_n, 0, \cdots, 0]$.
 Starting with our 3d GLSM, we couple it with a 1d $\mathcal{N}=2$ quiver gauge theory as shown in figure \ref{fig: groth-defect}. Each two consecutive nodes of the quiver are connected by a bifundamental chiral multiplet. In addition, at each node $U(\ell)$, we couple the 1d gauge theory with the $SU(n_f)$ 3d flavour symmetry via $M_\ell$ fermi multiplets $\Lambda^{(\ell)}_{\alpha^{(\ell)}}$ which are defined in terms of the partition $\mu$ as follows:
\begin{align}\label{para-number}
    \begin{split}
        &M_\ell = \mu_\ell - \mu_{\ell+1} +1 ~,\qquad \ell = 1, \cdots, n-1~,\\
        &M_n = \mu_n - n + N_c~.
    \end{split}
\end{align}
Moreover, for $\ell=1, \cdots, n$, we take the index $\alpha^{(\ell)}$ of the fermi field $\Lambda^{(\ell)}_{\alpha^{(\ell)}}$ to be an element of:
\begin{equation}\label{para-fermi-label}
  I_{\ell} \equiv \left\{1 + \sum_{k=\ell+1}^{n}M_k,~ 2+\sum_{k=\ell+1}^{n}M_k, \cdots,~ M_\ell + \sum_{k=\ell+1}^{n}M_{k}\right\}~.
\end{equation}
The 1d quiver is connected with the 3d matter fields $\phi$ via the 1d superpotential:
\begin{equation}\label{J-superpotential}
  L_J=   \int d\theta \sum_{\ell=1}^{n} \sum_{\alpha^{(\ell)}\in I_\ell}J^{(\ell)}_{\alpha^{(\ell)}}(\varphi; \phi) \Lambda^{(\ell)}_{\alpha^{(\ell)}}~,
\end{equation}
where, 
\begin{equation}
     J^{(\ell)}_{\alpha^{(\ell)}}(\varphi; \phi) \equiv \varphi_{\ell}^{\ell+1}\cdot \varphi_{\ell+1}^{\ell+2} \cdots \varphi_{n-1}^{n}\cdot\varphi_{n}^{n+1} \cdot \phi^{(\ell)}_{\alpha^{(\ell)}}~.
\end{equation}
Here, we decomposed the $N_c\times n_f$ matrix $\phi=(\phi^a_i)$ into blocks as follows:
\begin{equation}\label{block-decomp}
\phi = \large \left(
    \renewcommand{\arraystretch}{1.2}
    \begin{array}{c|c|c|c|c}
    \phi^{(n)}\; &\; \phi^{(n-1)}\;&\; \cdots\; &\; \phi^{(1)}\; & \;\phi^{(0)} \\
    \end{array}
\right)~,
\end{equation}
with $\phi^{(\ell)}$ is a block of size $N_c\times M_\ell$, and $\sum_\ell M_\ell = n_f$.

From the  perspective of the matrix $\phi$, this superpotential imposes the following linear constraints on the form of the 3d fields:
\begin{equation}
    \varphi_{\ell}^{\ell+1}\cdot \varphi_{\ell+1}^{\ell+2} \cdots \varphi_{n-1}^{n}\cdot\varphi_{n}^{n+1} \cdot \phi^{(\ell)}_{\alpha^{(\ell)}} = 0~,
\end{equation}
for $\ell=1, \cdots, n$ and $\alpha^{(\ell)}\in I_{\ell}$. Moreover, we have to solve the 1d D-term equations:
\begin{equation}\label{D-equations}
    \varphi_{\ell}^{\ell+1}\cdot \varphi_{\ell}^{\ell+1~\dagger} - \varphi_{\ell-1}^{\ell~\dagger} \cdot\varphi^{\ell}_{\ell-1} = \zeta_\ell~ \mathbb{I}_{\ell}~,\quad \qquad \ell = 1, \cdots , {n}~,\
\end{equation}
with $\zeta_\ell$ being the 1d real FI parameter associated with $U(1)\subset U(\ell)$ and here we take them to be positive. Solving  these equations recursively for the fields $\varphi$ and $\phi$, we can show that the final form of the matrix $\phi$ is precisely the one representing the Schubert cell C$_\mu$~\cite{Closset:2023bdr}. This means that the insertion of the line defect at a point on $\Sigma_g$ (and wrapping $S^1$) constrains the quasi-maps $\phi : \Sigma \rightarrow X$ to lie in the Schubert variety $X_\mu \subset X$.

\medskip
\noindent
\textbf{Grothendieck lines: computing the 1d flavoured index.} So far, we have shown that whatever coherent sheaf this half-BPS line $\mathcal{L}_\mu$ corresponds to must be supported on the (closure of the) Schubert cell C$_\mu$. Now, we compute the flavoured Witten index of the 1d $\CN=2$ supersymmetric quantum mechanics (SQM). By standard arguments, this 1d index must correspond to the Chern character of the coherent sheaf on $X$ that the line operator flows into. 

Following  well-established supersymmetric localization results~\cite{Hori:2014tda}, the Witten index of the 1d quiver introduced in figure \ref{fig: groth-defect} can be computed as a so-called JK residue:
\begin{equation}\label{cont-int-Z}
    \mathcal{L}_{\mu}(x,y) =  \oint_{\text{JK}} \left[\prod_{\ell=1}^{n}\frac{1}{\ell ! } \prod_{i_\ell=1}^{\ell} \frac{-d z_{i_\ell}^{(\ell)}}{2\pi i z_{i_\ell}^{(\ell)}} \prod_{1\leq i_\ell \neq j_\ell \leq \ell} \left(1-\frac{z_{i_\ell}^{(\ell)}}{z_{j_\ell}^{(\ell)}}\right)\right] \text{Z}^{\rm 1d}_{\text{matter}}(z,x,y)~,
\end{equation}
where $z_{i_\ell}^{(\ell)}$ are the gauge parameters of the 1d gauge group $U(\ell)$, and,  
\begin{equation}\label{Z-matter}
	\text{Z}^{\rm 1d}_{\text{matter}}(z,x,y) \equiv \prod_{\ell=1}^{n-1} \prod_{i_{\ell}=1}^{\ell} \frac{\prod_{\alpha^{(\ell)}\in I_\ell}\left(1-\frac{z_{i_\ell}^{(\ell)}}{y_{\alpha^{(\ell)}}}\right)}{\prod_{j_{\ell+1}=1}^{{\ell+1}}\left(1- \frac{z_{i_\ell}^{(\ell)}}{z_{j_{\ell+1}}^{(\ell+1)}}\right)} \prod_{i_n=1}^{n} \frac{\prod_{\alpha^{(n)}\in I_n} \left(1-\frac{z_{i_n}^{(n)}}{y_{\alpha^{(n)}}}\right)}{\prod_{a=1}^{N_c}\left(1-\frac{z_{i_n}^{(n)}}{x_a}\right)}~,
\end{equation}
is the contribution of the bifundamental and fermi multiplets appearing along the quiver. We must take all the 1d FI parameters $\zeta_\ell$ to be positive, in which case the JK residue prescription  instructs us to consider the contributions of the singularities coming from the matter multiplets only. Indeed, starting with the integral associated with 1d $U(1)$ gauge node and moving towards $U(n)$, while taking advantage of the residual Weyl symmetries on the Coulomb branch of the quantum mechanics, one finds that the above integral yields the following explicit expression:
\begin{equation}\label{Z-final-form}
	{\mathcal{L}}_{\mu}(x,y) = (-1)^{\bullet} \sum_{\mathcal{J}} \prod_{\ell=1}^{n} \left[\left(\prod_{j_{\ell}\in  \tilde{J}_{\ell}} x_{j_{\ell}}\right) \prod_{i_{\ell}\in J_\ell} \frac{\prod_{\alpha^{(\ell)}\in I_\ell}\left(1-x_{i_{\ell}} y_{\alpha^{(\ell)}}^{-1}\right)}{\prod_{j_{\ell+1}\in J_{\ell+1}\smallsetminus J_{\ell}}\left(x_{i_\ell} - x_{{j}_{\ell+1}}\right)}\right]~,
\end{equation}
where we sum over the set of subsets:
\begin{equation}\label{J-index-set}
	\mathcal{J} = \{J_1,\; J_2,\; \cdots,\; J_n\}~, 
\end{equation}
such that:
\begin{equation}
    J_1 \subset J_2 \subset \cdots \subset J_n \subseteq   \{1, \cdots, N_c\}~, \qquad |J_\ell| = \ell ~,
\end{equation}
and, $\tilde{J}_{\ell} \equiv \{1, \cdots, N_c\}\setminus J_{\ell}$.  One can then show that the twisted index of the SQM that we coupled to the 3d GLSM gives us the double Grothendieck polynomial $\mathfrak{G}_{\mu}(x,y)$ introduced in \eqref{doub-groth-poly}. Therefore, we can be confident that the half-BPS line operator constructed above indeed flows to the Schubert class $[\mathcal{O}_{\mu}]$ in the IR.

\begin{figure}[t]
    \centering
    \begin{tikzpicture}[baseline=1mm]
\node[node3d] (1) []{$N_c$};
\node[square,draw,minimum size=1.8cm,very thick] (2) [below= 2 of 1]{$n_f$};
\node[node1d] (rn) [left= 1.5 of 1]{$N_c$};
\node[node1d] (rn-1) [left= 1.5 of rn]{{$N_c-1$}};
\node[] (inter1) [left= 1.2 of rn-1]{$\;\cdots\;$};
\node[node1d] (r1) [left= 1.2 of r2]{$1$};
\draw[->-=0.6,very thick] (1) --  (2) node[midway,right]{$\; \; \phi_{\text{point/line}}$};
\draw[->-=0.6, thick] (rn) --  (1) node[midway,above]{$\varphi_{N_c}^{N_c+1}$};
\draw[->-=0.6, thick] (rn-1) --  (rn) node[midway,above]{$\varphi_{N_c-1}^{N_c}$};
\draw[->-=0.6, thick] (inter1) -- (rn-1) node[midway,above]{$\varphi_{N_c-1}^{N_c}$};
\draw[->-=0.6, thick] (r1) -- (inter1) node[midway,above]{$\varphi_1^{2}$};
\draw[->-=0.6,red, dashed, thick] (r1) -- (2) node[midway,below]{$\{n_f-1\}\;\;\;\;\;$};
\draw[->-=0.6,red, dashed, thick] (rn-1) -- (2) node[midway,above]{\;\;$I^{(p/l)}_{N_c-1}$};
\draw[->-=0.6,red, dashed, thick] (rn) -- (2) node[midway,above]{$\;\;\;\;I^{(p/l)}_{N_c}$};
  \node[fit=(r1)(rn), dashed, blue, draw, inner sep=10pt, minimum width=10.6cm,minimum height=5cm, shift={(0.6cm,-1.5cm)}] (rn) {};
\end{tikzpicture}
     \caption{Grothendieck defect associated with the point/line Schubert cell inside Gr$(N_c, n_f)$. Here the index sets differ between the two cases. For the point-case they are given by \eqref{point-para}, meanwhile, for the line-case, they are given by \eqref{line-para}.}
    \label{fig: point-groth-defect}
\end{figure}
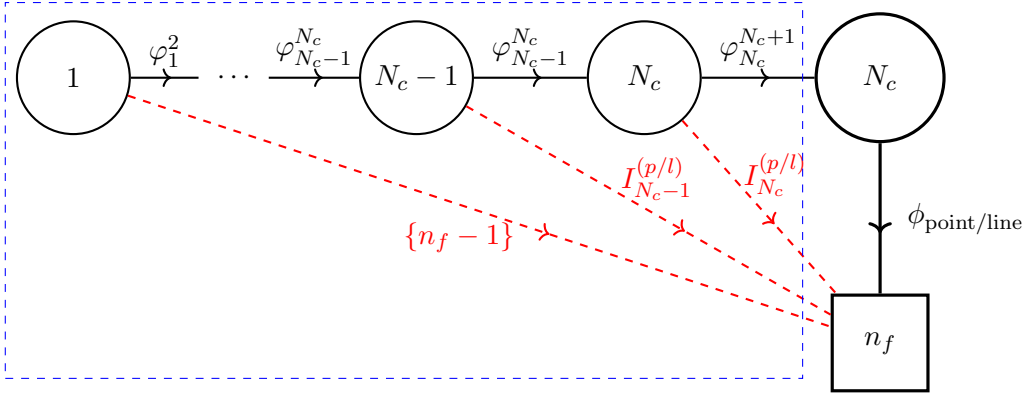

\medskip
\noindent
\textbf{Example: the point and the line.} As an example of the construction above, let us look at the following two special cases: the point and the line cells inside Gr$(N_c, n_f)$. In the first case, we have the partition:
\begin{equation}
    \mu_{\text{point}} = [n_f-N_c, \cdots, n_f-N_c]~.
\end{equation}
Following our labeling of fermi multiplets \eqref{para-fermi-label},
  the index sets for the fermi multiplets are:
\begin{equation}\label{point-para}
   I^{({p})}_{\ell} =\begin{cases}
       \{n_f-\ell\}~, \qquad &\ell = 1, \cdots, N_c-1~,\\
        \{1, \cdots, n_f-N_c\}~, \qquad &\ell = N_c~.
   \end{cases} 
\end{equation}
Moreover, one can check~\cite{Closset:2023bdr} that the final form of the matter matrix $\phi_{\text{point}}$ is:
\begin{equation}
    \phi_{\text{point}} = \begin{pmatrix}
        0_{N_c\times (n_f-N_c)} \mid 1_{N_c\times N_c}~
    \end{pmatrix}~,
\end{equation}
from which we deduce that, indeed, the dimension of the Schubert cell in this case is zero.

In the case of the line cell, we have the following $N_c$-partition:
\begin{equation}
    \mu_{\text{line}} = [n_f-N_c, \cdots n_f-N_c, n_f-N_c-1]~.
\end{equation}
In this case, the 1d fermi multiplets are indexing via:
\begin{align}\label{line-para}
        &I^{(l)}_{\ell} =\begin{cases}
            \{n_f-\ell\}~, \qquad &\ell = 1, \cdots, N_c-2~, \\
            \{n_f-N_c, n_f-N_c+1\}~, \qquad &\ell = N_c-1~, \\
            \{1, \cdots, n_f-N_c-1\}~, \qquad &\ell = N_c~.
        \end{cases} 
\end{align} 
The final form of the matter matrix $\phi_{\text{line}}$ in this case is given by:
\begin{equation}
   \phi_{\text{line}}  = \left(\begin{array}{c|c|c|c}
       0_{1\times (n_f-N_c-1)} &1&\star& 0_{1\times (N_c-1)} \\
       \hline
       0_{(N_c-1)\times (n_f-N_c-1)}  &0_{(N_c-1)\times 1} & 0_{(N_c-1)\times 1}& 1_{(N_c-1)\times (N_c-1)}  
   \end{array}\right)~,
\end{equation}
which shows that the Schubert cell in this case is one-dimensional.  In either case, the 1d quiver coupled to the 3d GLSM is given in figure \ref{fig: point-groth-defect} above.

\medskip
\noindent
\textbf{A 1d $\mathcal{N}=2$ duality move.}
The generic 1d quiver with gauge group as in figure~\ref{fig: point-groth-defect} can often be simplified using the following duality move~\cite{Closset:2023bdr}:
\begin{align}\label{duality move}
\begin{split}
\begin{tikzpicture}[baseline=3mm]
\node[](mid)  {};
\node[node1ds] (l3) [left = 4 of mid]{$r_{l+1}$};
\node[red] (M3) [below= 0.85 of l3]{$M_{l+1}$};
\node[node1ds] (l2) [left= 0.95 of l3]{$r_l$};
\node[red] (M2) [below= 0.85 of l2]{$M_{l}$};
\node[node1ds] (l1) [left= 0.95 of l2]{$r_{l-1}$};
\node[red] (M1) [below= 0.85 of l1]{$M_{l-1}$};
\draw[->-=0.6, thick] (l2) --  (l3) node[midway,above]{$\varphi_l^{l+1}$};
\draw[->-=0.6, thick] (l1) -- (l2) node[midway,above]{$\varphi_{l-1}^{l}$};
\draw[->-=0.6,red, dashed, thick] (l3) -- (M3);
\draw[->-=0.6,red, dashed, thick] (l2) -- (M2);
\draw[->-=0.6,red, dashed, thick] (l1) -- (M1);
\node[ ] (l3) [left = 3 of mid]{$\Rightarrow$};
 %
\node[node1ds] (r3) [right = -0.5 of mid]{$r_{l+1}$};
\node[red] (rM3) [below= 0.85 of r3]{$M_{l+1}$};
\node[node1ds] (r1) [left= 1.5 of r3]{$r_{l-1}$};
\node[red] (rM1) [below= 0.85 of r1]{$M_{l-1}{+}M_l$};
\draw[->-=0.6, thick] (r1) --  (r3) node[midway,above]{$\varphi_{l-1}^{l+1}$};
\draw[->-=0.6,red, dashed, thick] (r3) -- (rM3);
\draw[->-=0.6,red, dashed, thick] (r1) -- (rM1);
 \end{tikzpicture}
 \end{split}
\end{align}
which is applicable if and only if $r_l= r_{l+1}-M_l$.
 This is a special instance of a set of Seiberg-like dualities for 1d gauge theories similar to the 2d trialities of Gade, Gukov and Putrov~\cite{Gadde:2013lxa}, which will be discussed in future work. In the particular case when $r_{l-1}=0$, we can simply drop the $U(r_l)$ gauge node from the tail of the 1d quiver.

\begin{figure}[t]
    \centering
   \scalebox{0.75}{\begin{tikzpicture}[baseline=1mm]
  \node[node3d] (1l) []{$N_c$};
\node[square,draw,minimum size=1.8cm,very thick] (2l) [below= 2 of 1]{$n_f$};
\node[node1d] (rnl) [left= 1.5 of 1l]{$N_c$};
\node[node1d] (rn-1l) [left= 1.5 of rnl]{{$N_c-1$}};
\draw[->-=0.6,very thick] (1l) --  (2l) node[midway,right]{$\; \; \phi_{\text{line}}$};
\draw[->-=0.6, thick] (rnl) --  (1l) node[midway,above]{$\varphi_{N_c}^{N_c+1}$};
\draw[->-=0.6, thick] (rn-1l) --  (rnl) node[midway,above]{$\varphi_{N_c-1}^{N_c}$};
\draw[->-=0.6,red, dashed, thick] (rn-1l) -- (2l) node[midway,above]{\;\;$I^{(l)}_{N_c-1}$};
\draw[->-=0.6,red, dashed, thick] (rnl) -- (2l) node[midway,above]{$\;\;I^{(l)}_{N_c}$};
  \node[fit=(rn-1l)(rnl), dashed, blue, draw, inner sep=10pt, minimum width=6.3cm,minimum height=5cm, shift={(0.6cm,-1.5cm)}] (rnl) {};
\node[node3d] (1) [left = 4 of rn-1l]{$N_c$};
\node[square,draw,minimum size=1.8cm,very thick] (2) [below= 2 of 1]{$n_f$};
\node[node1d] (rn) [left= 1.5 of 1]{$N_c$};
\draw[->-=0.6,very thick] (1) --  (2) node[midway,right]{$\; \; \phi_{\text{point}}$};
\draw[->-=0.6, thick] (rn) --  (1) node[midway,above]{$\varphi_{N_c}^{N_c+1}$};
\draw[->-=0.6,red, dashed, thick] (rn) -- (2) node[midway,above]{$\;\;I^{(p)}_{N_c}$};
  \node[fit=(rn), dashed, blue, draw, inner sep=10pt, minimum width=3cm,minimum height=5cm, shift={(0.7cm,-1.5cm)}] (rn) {};
\end{tikzpicture}}
     \caption{Simplified form of the Grothendieck defects associated with the point (left quiver) and line (right quiver) Schubert cells in Gr$(N_c, n_f)$. These are obtained by applying the 1d triality \eqref{duality move} to the quivers in figure \ref{fig: point-groth-defect}. The index sets are given in \eqref{simp-p-ind-set} for the point and in \eqref{simp-l-ind-set} for the line.}
    \label{fig: simp-point-groth-defect}.
\end{figure}
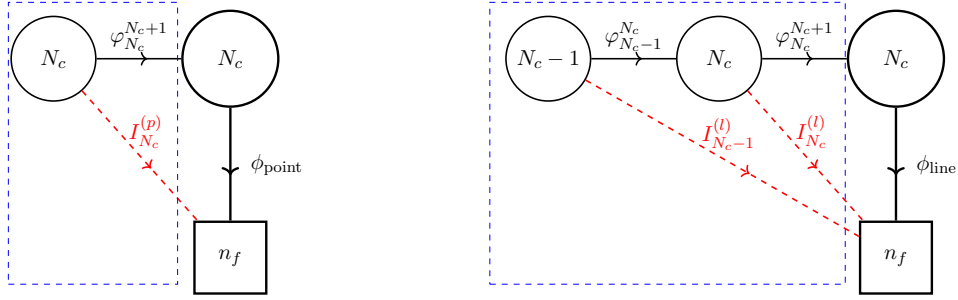

As an application of this duality move, we can simplify the form of the 1d quiver defining the point and line cells in figure~\ref{fig: point-groth-defect}. Starting from the left of the quiver, for instance, we can simply drop all the nodes which have a single fermi multiplet, to obtain the quivers in figure~\ref{fig: simp-point-groth-defect}. For the point-cell, we have the single index set:
\begin{equation}\label{simp-p-ind-set}
    I^{(p)}_{N_c} =    \{1, \cdots, n_f-N_c\}~, 
\end{equation}
and for the line-cell, we have the two index:
\begin{equation}\label{simp-l-ind-set}
    I^{(l)}_{\ell} = \begin{cases}
        \{n_f-N_c, n_f-N_c+1\}~, \qquad &\ell = N_c-1~, \\
        \{1, \cdots, n_f-N_c-1\}~,\qquad &\ell = N_c~.
    \end{cases}
\end{equation}

\medskip
\noindent 
\textbf{The 2d/0d limit of our constructioon.} One can follow the same construction as above for the 2d GLSM into $X$, where the point defect is now defined in terms of a $0$d $\mathcal{N}  = 2 $ quiver gauge theory with the same data that we chose for the 1d case. Computing the partition function of this supersymmetric matrix model gives us the double Schubert polynomial $\mathfrak{S}_{\mu}(\sigma, m)$, which is known to represent the equivariant Schubert class $[\text{C}_\mu]\in \text{H}^{2|\mu|}_{T}(\text{Gr}(N_c, n_f), \mathbb{Z})$. This polynomial has the following explicit form:
\begin{equation}\label{double-schubert-def}
    \mathfrak{S}_{\mu}(\sigma, m) \equiv \frac{\det_{1\leq a,b\leq N_c}\left(\prod_{\alpha=1}^{\mu_a +N_c - b}\left(\sigma_b - m_{\alpha}\right)\right)}{\prod_{1\leq b< a\leq N_c} \left(\sigma_a - \sigma_b\right)}~.
\end{equation}
One can take the small-circle limit, $\beta\rightarrow 0$, of the 3d discussion above, to retrieve the data of the QH$^{\bullet}(\text{Gr}(N_c, n_f))$ from the quantum K-theory ring. Indeed, in this limit, one finds the following relation between the double Grothendieck polynomials \eqref{doub-groth-poly} and the double Schubert polynomials \eqref{double-schubert-def}:
\begin{equation}
  \mathfrak{G}_{\lambda}(x, y)  \rightarrow     (2\pi\beta)^{|\lambda|}\, \mathfrak{S}_{\lambda}(\sigma, m) ~.
\end{equation}
Our 0d/2d construction thus provides a clear physical description of a standard basis for the equivariant quantum cohomology ring of $X$. (In the non-equivariant case, the Schubert polynomials reduce to Schur polynomials.)

\subsection{Quantum K-theory from the A-model}
In subsection \ref{subsec:frobalg} above, we discussed how one can compute the data of the twisted chiral algebra $\mathcal{R}^{\text{3d}}$ using computational algebraic geometry. In this section, we provide a more explicit formulas for K-theoretic enumerative invariants, which simply follow from the JK residue prescription~\cite{Benini:2015noa} for the 3d twisted index of the SQCD$[N_c,k, l, n_f,0]$ theories in the geometric window.

From supersymmetric localization, the path integral of the 3d twisted index on the sphere reduces to the following JK residue:
  \begin{equation}\label{JK-residue-formula}
    \left<\mathcal{L}\right>_{\mathbb{P}^1\times S^1}   = \frac{1}{N_c!} \sum_{\mathfrak{m}\in \mathbb{Z}^{N_c}} \sum_{x_* \in \tilde{\mathfrak{M}}_{\text{sing}}^{\mathfrak{m}}} \underset{x = x_*}{\text{JK-Res}} \left[\textbf{Q}(x_*), \eta_\xi \right]\mathfrak{I}_{\mathfrak{m}}[\mathcal{L}](x, y, q)~.
    \end{equation}
    This formula involves a  sum over all $U(N_c)$ magnetic fluxes $\mathfrak{m}=(\mathfrak{m}_a)\in \mathbb{Z}^{N_c}$ through $\mathbb{P}^1$, and a sum over JK residues in each flux sector.
 The latter are taken with respect to the $N_c$-form:
    \begin{equation}\label{JK form}
        \mathfrak{I}_\mathfrak{m}[\mathcal{L}](x, y, q)  = (-2\pi i)^{N_c} Z_{\mathfrak{m}}(x, y, q) \mathcal{L}(x, y, q)\; \frac{dx_1}{x_1} \wedge \cdots \wedge \frac{dx_{N_c}}{x_{N_c}}~,
\end{equation}
which has codimension-$N_c$ singularities (including at infinity) denoted by $\tilde{\mathfrak{M}}_{\text{sing}}$. For each magnetic  flux $\mathfrak{m}\in \mathbb{Z}^{N_c}$, the factor $Z_{\mathfrak{m}}$ is given in terms  of the effective dilaton potential and gauge flux operators as:
\begin{equation}
Z_{\mathfrak{m}}(x, y, q) =e^{-2\pi i \Omega} \prod_{a=1}^{N_c} \Pi_a(x, y, q)^{\mathfrak{m}_a}~, \qquad \Pi_a(x,y,q) \equiv e^{2\pi i \frac{\partial\mathcal{W}}{\partial u_a}}~.
\end{equation}
To each singularity $x_\ast\in \tilde{\mathfrak{M}}_{\text{sing}}$, one assigns a charge vector ${\textbf{Q}}(x_\ast)$ which determines whether or not the singularity contributes non-trivially to the JK residue, given a choice of the auxiliary parameter $\eta_\xi$. For the particualr choice $\eta_\xi = (\xi, \cdots, \xi)\in \mathbb{C}^{N_c}$, one can show that the sum over singularities $x_*$  is closely related to the sum over 3d vacua which we reviewed in section \ref{sec:mod-space-vacaua} above~\cite{Closset:2023bdr}.  

For example, for $\xi>0$ and CS levels $k$ and $l$ in the geometric window, we find that the residue \eqref{JK-residue-formula} receives contributions only from the singularities coming from the matter fields. Therefore, the formula \eqref{JK-residue-formula} in this case can be simplified further into the following form:
\begin{equation}\label{KGW}
\left<\mathcal{L}\right>_{\mathbb{P}^1\times S^1}  (q,y) = \sum_{d= 0}^\infty  q^{d} \;\textbf{I}_{d}[{\mathcal{L}}](y)~, 
   \end{equation}
where $d= |\mathfrak{m}|\equiv \sum_{a=1}^{N_c} \mathfrak{m}$ is the magnetic flux for the overall $U(1)\subset U(N_c)$. It corresponds to the degree of the holomorphic map $\phi\, : \, \Sigma \rightarrow X$ in the infrared NLSM realisation. 
 At each degree $d$, the residue $\textbf{I}_{d} [{\mathcal{L}}](y)$ has the explicit form:
\begin{equation}\label{Id explicit}
\sum_{\substack{\mathfrak{m}_a \geq 0 \\ |\mathfrak{m}|=d}}\;\; \sum_{1\leq \alpha_1<\cdots <\alpha_{N_c} \leq n_f} \underset{\{x_a = y_{\alpha_a}\}}{\text{Res}}\;\frac{(-1)^{|\mathfrak{m}|(K+l) +N_c}\Delta(x) {\mathcal{L}}(x,y)}{\prod_{a=1}^{N_c}x_a^{\mathfrak{r}_a}\prod_{\alpha=1}^{n_f}\left(1-x_a y_\alpha^{-1}\right)^{1+\mathfrak{m}_a}}~.
\end{equation}
Here, we defined $K= k+\frac{n_f}{2}$ and we took the $R$-charge $r=0$, which is the natural choice from the GLSM point of view. We also defined the integers:%
\begin{equation}\label{ra}
    \mathfrak{r}_a \equiv N_c  - l |\mathfrak{m}| - K \mathfrak{m}_{a}~, \qquad a = 1, \cdots, N_c~,
\end{equation}
at fixed $\mathfrak{m}$. Moreover, we have the Vandermonde determinant:
\begin{equation}\label{vandermonde}
    \Delta(x) \equiv \prod_{1\leq a\neq b\leq N_c} (x_a - x_b)~.
\end{equation}
Using this residue formula, we can compute the topological metric and the structure constants of QK$_T(\text{Gr}(N_c, n_f))$ in any basis, once we know how (equivariant) K-theory classes are represented by polynomials in $x$ (and $y$). Here we consider the insertion of the Grothendieck lines $\mathcal{L}_{\mu}$. For example, to compute the topological metric $\eta_{\mu\nu}$ we have:
\begin{equation}
    \eta_{\mu}(q,y) = \left<\mathcal{L}_{\mu}\mathcal{L}_\nu\right>_{\mathbb{P}^1\times S^1} = \sum_{d=0}^{\infty} q^d \textbf{I}_{d} \left[\mathfrak{G}_{\mu}\mathfrak{G}_\nu\right]~.
\end{equation}
Similarly, for the stucture constants, we have:
\begin{equation}
    \mathcal{N}_{\mu\nu\lambda} (q,y) = \left<\mathcal{L}_\mu\mathcal{L}_{\nu}\mathcal{L}_\lambda\right>_{\mathbb{P}^1\times S^1}=\sum_{d=0}^{\infty} q^d \textbf{I}_d\left[\mathfrak{G}_{\mu}\mathfrak{G}_{\nu}\mathfrak{G}_{\lambda}\right]~.
\end{equation}
In this way,  for example, we can check the relations \eqref{QKP2}, which were previously computed using the Gr\"obner basis techniques.

\section{Conclusions}
In this note, we reviewed recent results we obtained for the 3d $\CN=2$ unitary SQCD theory, especially in the case of the $U(N_c)$ theory with $n_f$ fundamental chiral multiplets and no antifundamental multiplets ($n_a=0$). We focused on the structure of its infrared vacua in the case with non-zero FI parameter. Using this infrared approach, in Ref~\cite{Closset:2023jiq}, we also obtained the flavoured Witten index for SQCD$[N_c, k, l, n_f, n_a]$ -- previously, it had only been computed for $l=0$. Here, we explained how the 3d GLSM approach (that is, the 3d $\CN=2$ theory compactified on a circle) leads to an elegant physics computation of the quantum K-theory of the Grassmannian manifold Gr$(N_c, n_f)$. We proposed a new set of half-BPS line defects, dubbed Grothendieck lines, which are defined as 1d/3d coupled systems in the UV gauge theory and flow to Schubert classes in the quantum K-theory. This provided a clean physical construction of very important mathematical objects. In particular, the Grothendieck polynomials that represent Chern characters of the Schubert classes arise as 1d Witten indices of linear quivers.

The natural next step would be to properly understand the quantum K-theory of  flag  varieties in the 3d GLSM language. These varieties are realised in terms of more general 3d $\CN=2$ linear quivers, and there appears to be a beautiful correspondence between the combinatorics of Schubert varieties in flag manifolds and the kinds of 1d/3d defects considered here. We hope to report on progress on this front in the near future.

\subsection*{Acknowledgements}  We are grateful to Mathew Bullimore,   Hans Jockers, Heeyeon Kim,  Leonardo Mihalcea, and Eric Sharpe for discussions and correspondence. 
CC is a Royal Society University Research Fellow supported by the University Research Fellowship Renewal 2022 `Singularities, supersymmetry and SQFT invariants'. The work of OK is supported by the School of Mathematics at the University of Birmingham.


\bibliographystyle{JHEP}
\bibliography{3dbib}

\end{document}